# Bone Cancer Rates in Dinosaurs Compared with Modern Vertebrates

L.C. Natarajan[1], A.L. Melott[1], B.M. Rothschild[2], and L.D. Martin[2]

1 Department of Physics and Astronomy, University of Kansas, Lawrence, KS 66045
2 Biodiversity Research Center, University of Kansas, Lawrence, KS 66045

## ABSTRACT

Data on the prevalence of bone cancer in dinosaurs is available from past radiological examination of preserved bones. We statistically test this data for consistency with rates extrapolated from information on bone cancer in modern vertebrates, and find that there is no evidence of a different rate. Thus, this test provides no support for a possible role of ionizing radiation in the K-T extinction event.

## INTRODUCTION

Cancer is one of the few diseases that can be reliably diagnosed in vertebrate remains (Rothschild and Martin 2006). Cancer prevalence may potentially help shed light on biosphere stress. Some scenarios for biodiversity fluctuation and mass extinctions include radiation effects, which might increase the cancer rate (e.g. Medvedev and Melott 2007). Since dinosaurs were involved in a major mass extinction, it is interesting to see whether there is any evidence of an elevated cancer rate.

While primary bone cancer can be defined as cancer that begins in the bone, secondary or metastatic bone cancer is cancer that originates elsewhere in the body and spreads to the bone. Bone is the third most common site for metastasis after lung and liver (Alvarez 1948; Abrams, Spiro and Goldstein 1950; Wierman, Clagett and McDonald 1954; Resnick 2002). Whereas a variety of cancers can spread to the bone, the most common forms in humans are breast, prostate, lung, kidney and thyroid (American Cancer Society 2007). In this study, the prevalence of secondary bone cancer in dinosaurs is examined for consistency with rates in modern vertebrates.

While 2-3% benign hemangiomas (proliferation of vascular endothelium) were found in the family *Hadrosauridae*, only one case of metastatic cancer was found (within the same family). The diseased vertebra was that of an *Edmontosaurus* (Rothschild et al. 2003) specimen from the Maastrichtian stage. It is particularly interesting that this case came from near the end-Cretaceous, when the extinction event took place. We note that the elevated rate of hemangiomas was also Maastrichtian, which is interesting, but we have not studied this question since it cannot be excluded that, for example, it is a consequence of the diet or some other peculiarity of this family. The focus of our study is to do a statistical test for consistency between the rates in dinosaurs versus modern vertebrates.

## DATA AND METHODS

In order to test for consistency, we need to establish an expected rate based on data from modern vertebrates, and then ask whether the event in the dinosaur fossil record is consistent with that rate.

While the spine is the most commonly affected area of bone metastasis in the human skeleton, the next most common areas of spread include the ilium, pubis, ischium, proximal femur, femora and skull (American Cancer Society 2007). Thus, if a more comprehensive study is to be performed, examination of these regions is suggested. Furthermore, as suggested by Rothschild and Rothschild (1995), visual examination is complementary to radiologic examination and therefore should be included if the epidemiology of metastatic disease is to be determined reliably. However, the radiologic method alone was the basis of the sample and control rate estimates used in this study.

The prevalence of bone cancer in human skeletons was examined. We justify the comparison based on a superficial similarity of large vertebrates with substantial life spans (Erickson et al. 2006). The rate of bone metastasis was estimated with respect to x-ray identified metastatic disease in the Hamann-Todd Collection (Rothschild and Rothschild 1995).

The Hamann-Todd Collection is the largest and best preserved compilation of human skeletons for which a background demographic is known (Rothschild and Woods 1991; Rothschild and Martin 1993). From a total of 2906 defleshed skeletons, 33 cases of metastatic disease were identified fluoroscopically, yielding a probability of 1.14 %. Based on necropsy results of captive wild animals (Effron, Griner and Benirschke 1977), the rate of cancer in reptiles is approximately $1/8^{th}$ that found in humans. Thus, the rate of cancer in dinosaurs can be tested for consistency with a rate of 0.142%.

Another estimate can be made based on the rate in birds. The rate of macroscopically observable cancer in birds was less than 24 in 50,000 (Rothschild and Panza 2005), indicating that the rate of cancer in dinosaurs can be tested for consistency with 0.048%. The Poisson distribution was subsequently applied to both rates in order to calculate probability. Our null hypothesis is that the rate of such cancer in dinosaurs is not higher than either of these rates.

The data concerning the prevalence of cancer in the fossil record was presented in Rothschild et al. (2003). The epidemiology of tumors in dinosaurs was investigated by fluoroscopically screening dinosaur vertebrae, a technique which allowed the examination of vertebrae in real time, thus negating the need for film. A total of 10,312 vertebrae from 708 individual dinosaurs of varying families were examined, and one such tumor was found.

**RESULTS**

The malignant *Edmontosaurus* tumor reported by Rothschild et al. (2003) is metastatic cancer of unknown (primary) origin. In the table below, we show the probability of finding various numbers of subjects with tumors (left column) in a sample of 708 examined, assuming the rate for reptiles (center column) and birds (right column) It was determined that the rate of metastatic bone disease in the dinosaur fossil record is in fact consistent with the predicted rate of bone metastasis in the dinosaur record (Table 1). The probability of finding less than one cancer is 37% according to the reptilian cancer rate, and 71% according to the avian rate, so the observed event does not represent an elevated rate. We cannot negate the null hypothesis. Thus, there does not appear to be an elevated rate of bone cancer in the dinosaur fossil record. Note that if the null hypothesis had been formulated as "not lower rate", we would not be able to reject that either, based on the size of the sample and the expected rates.

| Dinosaurs (708) | Reptiles | Birds |
|---|---|---|
| # with cancer | Probability | Probability |
| 0 | 0.36605 | 0.71188 |
| 1 | 0.36787 | 0.24193 |
| 2 | 0.18486 | 0.04111 |
| 3 | 0.06193 | 0.00466 |
| 4 | 0.01556 | 0.00040 |
| 5 | 0.00313 | 0.00003 |
| 6 | 0.00052 | 0.00000 |
| 7 | 0.00008 | 0.00000 |
| 8 | 0.00001 | 0.00000 |

**Table 1** Examination of vertebrae from 708 dinosaurs revealed 1 case of metastatic disease. Thus, the frequency of metastatic bone cancer is consistent with the risk estimated from rates in modern vertebrates, as determined using the Poisson distribution.

## DISCUSSION

We have asked whether the bone cancer rate in dinosaurs is consistent with rates for modern vertebrates. Results of past examination of dinosaur fossil bones for evidence of metastatic bone cancer are tested for consistency with available information on rates in modern vertebrates. We were able to make two rate estimates, not more than a factor of three apart, based on available data. We then ask about the probability of finding one bone cancer in 708 dinosaurs based on these assumed rates. We find that with either rate, this is not an unexpected outcome. We conclude that there is no evidence for an abnormal rate of bone cancer in dinosaurs.

Exposure to ionizing radiation can elevate the rate of bone cancers (Brenner et al. 2003). A variety of astrophysical sources of such radiation have been hypothesized to have a role in extinction events (e.g. Fields 2004; Medvedev and Melott 2007). It is therefore interesting to ask whether there is evidence for an elevated cancer rate as a possible signal for a role in ionizing radiation. The answer is no. There is however, a residual question of a somewhat elevated rate of benign hemangiomas in the family *Hadrosauridae* which is outside the scope of this study, but may deserve further consideration in the future.

## ACKNOWLEDGEMENTS

This research was supported by NASA Astrobiology: Exobiology and Evolutionary Biology grant NNG04GM14G, the University of Kansas Honors Program, and by the KU Student Senate.